\documentclass{PoS}

\usepackage{slashed}

\def\wp{\widetilde P}

\def\beq{\begin{equation}}
\def\eeq{\end{equation}}
\def\beeq{\begin{eqnarray}}
\def\eeeq{\end{eqnarray}}
\def\cm{{\cal M}}

\def\to{\rightarrow}

\def\nn{\nonumber}

\def\s#1#2{s_{#1#2}}

\def\ID{1 \kern -.45 em 1}

\def\vep{{\varepsilon}}

\def\b0{b_0}

\def\s{{\rm Split}}

\title{Recursion relations for the multiparton collinear limit and
splitting functions}

\ShortTitle{Recursion relations for the multiparton collinear limit and
splitting functions}

\author{Stefano Catani\\
       INFN, Sezione di Firenze and Dipartimento di Fisica e Astronomia,\\ 
       Universit\`a di Firenze, I-50019 Sesto Fiorentino, Florence, Italy\\
       E-mail: \email{catani@fi.infn.it}}
\author{\speaker{Petros Draggiotis} \\
       Instituto de F\'{\i}sica Corpuscular, 
       Consejo Superior de Investigaciones Cient\'{\i}ficas-Universitat de Val\`encia, 
       Parc Cient\'{\i}fic, E-46980 Paterna, Valencia, Spain\\
       E-mail: \email{Petros.Drangiotis@ific.uv.es}}
\author{German Rodrigo\\
       Instituto de F\'{\i}sica Corpuscular, 
       Consejo Superior de Investigaciones Cient\'{\i}ficas-Universitat de Val\`encia, 
       Parc Cient\'{\i}fic, E-46980 Paterna, Valencia, Spain\\
       E-mail: \email{german.rodrigo@csic.es}}

\abstract{We present a systematic method to evaluate the splitting functions for tree-level QCD processes where $m$ partons
          approach the collinear limit. The splitting functions are computed by deriving on-shell recursion equations,
          which are similar to the Berends--Giele recursion relations for off-shell currents.}

\FullConference{``Loops and Legs in Quantum Field Theory '' 11th DESY Workshop on Elementary Particle Physics \\
                April 15-20, 2012\\
                Wernigerode, Germany}

\begin{document}

\section{Introduction}

One of the main features of perturbative scattering amplitudes in QCD and, more generally, gauge field theories
is the presence of singularities in the infrared (soft and collinear) regions of the phase space.
The knowledge of this singular behaviour is very relevant to make reliable QCD predictions through
high-order perturbative computations, all-order resummed calculations and parton-shower
Monte Carlo generators.

In this contribution we deal with the collinear limit and the associated singular behaviour
\cite{Altarelli:1977zs}--\cite{Forshaw:2012bi}.
We refer to a generic scattering amplitude in the kinematical configuration where the momenta of 
$m \;(m \geq 2)$ external QCD partons become parallel. In this multiparton collinear limit, the scattering 
amplitude fulfils a factorization formula: the factor that captures the singular collinear behaviour is 
a `splitting function' that is universal (process independent). The splitting function, which can be
presented and computed either in a colour-stripped form (the {\em splitting amplitude})
\cite{Berends:1987me, Mangano:1990by}
or in a colour-dressed form (the {\em splitting matrix}) 
\cite{Catani:2003vu},
effectively describes the collinear splitting subprocess 1~parton~$\to$~$m$~partons.
Applications to fixed-order calculations at the 
Next-to-Next-to-Leading Order
(NNLO) and to resummed calculations or parton-shower algorithms
at the Next-to-Next-to-Leading Logarithmic (NNLL) accuracy require the {\em known}
splitting functions for the one-loop $1 \to 2$
\cite{Bern:1998sc, Kosower:1999rx}
and the tree-level $1 \to 3$ 
\cite{Campbell:1997hg, Catani:1998nv, DelDuca:1999ha}
splitting subprocesses. The multiparton splitting subprocesses $1 \to m$ with higher multiplicity
$( m \geq 4)$ enter calculations at still higher orders.

In this talk we consider the multiparton collinear limit at the {\em tree-level}. The explicit computations
of the tree-level splitting functions with $m \leq 3$ partons 
\cite{Altarelli:1977zs}--\cite{DelDuca:1999ha}
have been carried out with methods and techniques that can also be extended and applied to the cases
with $m \geq 4$. However, these extensions are certainly cumbersome in practical terms, especially
if the number $m$ of collinear partons increases. Therefore, more practical methods
are desirable. The authors of 
Ref.~\cite{Birthwright:2005ak}
have used the MHV rules 
\cite{Cachazo:2004kj}
(they have also investigated the use of the BCFW recursion relations 
\cite{Britto:2004ap})
to compute multiparton splitting amplitudes: considering some specific classes of helicity configurations
of the collinear partons, these authors have derived general results that are valid for an arbitrary
number $m$ of gluons plus up to four fermions.

We have developed an alternative method 
\cite{Catani:2012tb}
to compute the tree-level
splitting functions for the multiparton collinear limit of gluons, quarks and antiquarks.
The method leads to recursion relations that apply directly to the splitting functions.
Starting from the splitting functions for $m=2$ and $m=3$ collinear partons, the recursion equation
iteratively gives the splitting functions for an arbitrary number of collinear partons.
For simplicity, in the following sections we illustrate the recursion relations for the pure gluon case.

\section{The multiparton collinear limit and factorization}

We consider a generic (on-shell) scattering amplitude $\cm(p_1,p_2,\dots)$
at the tree level. The momenta of the
external QCD partons are $p_1, p_2$ and so forth.
Throughout this presentation we use the notation
$p_{i,j}=p_i+p_{i+1}+\ldots +p_j$ and  $s_{i,j}=(p_i+p_{i+1}+\ldots +p_j)^2$,
with $i < j$.

The collinear limit of a set $\{ p_1, \dots, p_m \}$ of
$m$ ($m \geq 2$) parton momenta is approached when the momenta
of the $m$ partons become parallel.
This implies that all the parton subenergies
\beq
s_{i \ell}=(p_i+p_\ell)^2 \;\;,
\quad \quad {\rm with} \;\;\;\;\; i,\ell \in \{\,1,\dots,m \,\} \;\;, 
\eeq
are of the {\em same} order and vanish {\em simultaneously}
\cite{Campbell:1997hg,Catani:1998nv}.
To specify the kinematics of the $m$-parton collinear limit,
we define the light-like momentum
$\wp_{1,m}^\mu$:
\beq
\label{ptilm}
{\wp}_{1,m}^\mu \equiv 
p_{1,m}^\mu   
- \frac{p_{1,m}^2 }{2 \, n \cdot p_{1,m}} \; n^\mu \;\;, 
\eeq
where $n^\mu$ is an auxiliary light-like vector ($n^2=0$), which parametrizes how the 
collinear direction is approached.
In the multiparton collinear limit
we have
$p_i^\mu \to z_i {\wp}_{1,m}^\mu$ ($i=1,\dots,m$), and the longitudinal-momentum 
fraction $z_i$ is
\beq
\label{zim}
z_i = \frac{n \cdot p_i}{n \cdot {\wp}_{1,m}} =
\frac{n \cdot p_i}{n \cdot (p_1 + \dots + p_m)} \;\;. 
\eeq

In the following we limit ourselves to considering pure multigluon amplitudes. 
The $n$-gluon scattering amplitude is 
$\cm^{a_1,a_2,\dots,a_n}(p_1,p_2,\dots,p_n)$ and $a_1,a_2,\dots,a_n$ are the colour indices 
of the gluons. The scattering amplitude
$\cm^{a_1,a_2,\dots,a_n}$ can be decomposed in colour subamplitudes
\cite{Berends:1987me, Mangano:1990by}.
The colour-ordered (and colourless) subamplitude is denoted by 
$A_n(i_1,\dots,i_n)$, and the argument $i_k$ ($i_k \in \{1,\dots,n\}$) denotes the 
dependence on the $i_k$-th gluon, i.e. on its {\em outgoing} momentum $p_{i_k}^\mu$
and its polarization vector $\vep^\nu(p_{i_k})$ (the helicity states of $\vep^\nu$
are never explicitly denoted throughout the present contribution).

In the $m$-gluon collinear limit, the colour-ordered amplitude $A_n$ (with
$n \geq m+3$) fulfils the following {\em tree-level} factorization formula
\cite{Berends:1987me, Mangano:1990by, Campbell:1997hg,Catani:1998nv}:
\beq
\label{facttreesub}
A_n(\dots,k,1,2,\dots,m,j,\dots) \simeq \,{\s}(1,2,\dots,m;\wp_{1,m}) 
\;\;\;A_{n+1-m}(\dots,k,\wp_{1,m},j,\dots) \;\;,
\eeq
where the {\em splitting amplitude}
${\s}(1,2,\ldots,m;\wp_{1,m})$
has the singular behaviour
${\s} \propto (1/{\sqrt {s_{1,m}}})^{m-1}$,
and the neglected terms on the right-hand side are less singular in the collinear limit.

The splitting amplitude ${\s}(1,2,\ldots,m;\wp_{1,m})$
is universal (e.g., it is independent of $A_n$) and its depends on the collinear gluons 
and on the parent collinear gluon of the splitting subprocess $1$~gluon~$\to m$~gluons.
The parent gluon has {\em ingoing} momentum $\wp_{1,m}^\mu$ and polarization vector
$\vep^*_\nu(\wp_{1,m})$ ($\vep^*_\nu$ is the complex conjugate of $\vep_\nu$). 
Note that the product ${\s}(\dots;\wp_{1,m}) \;A_{n+1-m}(\dots,\wp_{1,m},\dots)$
involves a sum (which is not explicitly denoted on the right-hand side of 
Eq.~(\ref{facttreesub})) over the polarization states of the parent collinear gluon.
Thus, $\s$ has to be formally regarded as a matrix in the spin polarization (helicity)
space of the gluons.

The splitting amplitude ${\s}(1,2,\ldots,m;\wp_{1,m})$ is an {\em on-shell}
quantity and it is colour-ordered (analogously to $A_n$) with respect to the $m$ collinear 
gluons. Note also that, on the left-hand side of Eq.~(\ref{facttreesub}), the gluon
indices $1,\dots,m$ in the argument of $A_n$ are adjacent. If these indeces are not 
adjacent, the corresponding amplitude $A_n$ is subdominant in the $m$-gluon collinear limit.

We recall that the all-loop amplitude fulfils a factorization formula that is {\em partly}
similar to the tree-level formula in Eq.~(\ref{facttreesub}).
If the multiparton collinear limit occurs in the {\em time-like} region, the factorization 
formula 
\cite{Kosower:1999xi}
is exactly analogous to Eq.~(\ref{facttreesub}).
If instead the collinear limit occurs in the {\em space-like} region, the universality
structure of collinear factorization is violated 
\cite{Catani:2011st}, 
and the corresponding loop splitting amplitude acquires an explicit process dependence
(i.e., $\s$ depends on the adjacent {\em non-collinear} legs $k$ anf $j$ of $A_n$
in Eq.~(\ref{facttreesub}) at one-loop order, and it depends on additional adjacent
{\em non-collinear} gluons at higher-loop orders
\cite{Catani:2011st}).

\section{The recursion relation for the multigluon splitting amplitude}
\label{sec:recrel}

The splitting amplitude $\s(1,\ldots,m;\wp_{1,m})$ of $m$ gluons can be directly expressed 
and computed in terms of the corresponding splitting amplitudes of a smaller number 
$k$ ($k < m$) of gluons. This iterative structure follows from recursion relations that 
are derived in 
Ref.~\cite{Catani:2012tb}
for the general multiparton collinear limit of gluons, quarks and antiquarks.

The recursion relation for the multigluon splitting amplitude is 
\cite{Catani:2012tb}
\beeq
\s(1,\ldots,m;\wp_{1,m}) = \frac{1}{s_{1,m}}&&\!\!\!\!\! \left[ \, \sum_{k=1}^{m-1}
 \s(1,.., k;\wp_{1,k})   \right.  
\,  \s(k+1,.., m;\widetilde P_{k+1,m})\; 
V^{(3)}\!(\widetilde P_{1,k}, \widetilde P_{k+1,m};\wp_{1,m}) \nn \\
&&+  \, 
\sum_{k=1}^{m-2}\;  \sum_{l=k+1}^{m-1}
\s(1, \ldots, k;{\widetilde P_{1,k}}) \;\;  
\s(k+1, \ldots, l;{\widetilde P_{k+1,l}}) \
\nn \\ &&\times \left. \;
\s(l+1, \ldots, m;{\widetilde P_{l+1,m}}) 
\;\; V^{(4)}\!(\widetilde P_{1,k}, \widetilde P_{k+1,l}, \widetilde P_{l+1,m};{\wp}_{1,m})
\, \right]~,
\label{eq:generalgluon}
\eeeq
where, on the right-hand side, the splitting amplitude of a single gluon is 
$\s(i;\wp)=1$ by definition.
We recall that the function $\s$ depends on the polarization (helicity) states of the parent
collinear gluon. Therefore, the right-hand side of Eq.~(\ref{eq:generalgluon})
involves sums (which are not explicitly denoted) over the polarization states of the parent collinear gluons with momenta  
$\wp_{1,k}, \widetilde P_{k+1,m}, \widetilde P_{k+1,l}$ and 
$\widetilde P_{l+1,m}$.

The factors  $V^{(3)}$ and $V^{(4)}$ are a three-gluon and a four-gluon effective vertex, 
respectively.
The explicit expressions of the effective verices are
\beeq
\label{colord3g}
V^{(3)}\!(\widetilde P_1, \widetilde P_2;\wp) = g_S \;\frac{1}{\sqrt 2}
&& \left[ \, 
\vep(\widetilde P_1) \cdot \vep(\widetilde P_2) 
\;\left(\widetilde P_1 -\widetilde P_2 \right) \cdot \vep^{*}(\wp) \right. \nn \\
&&+ \left. \vep(\widetilde P_2) \cdot \vep^{*}(\wp)  
\;2 \,\widetilde P_2 \cdot \vep(\widetilde P_1) 
- \vep(\widetilde P_1) \cdot \vep^{*}(\wp)  
\;2 \,\widetilde P_1 \cdot \vep(\widetilde P_2) 
\right] \;\;,
\eeeq
\beeq
\label{colord4g}
V^{(4)}\!(\widetilde P_1, \widetilde P_2, \widetilde P_3 ;\wp) = g_S^2 &&
\Bigl\{ \;\;
 \vep(\widetilde P_1) \cdot \vep(\widetilde P_3) \;\; \vep(\widetilde P_2) \cdot \vep^{*}(\wp)
\Bigr. \nn \\
&&
 + \frac{n \cdot \widetilde P_1  \; n \cdot \widetilde P_2 
- n \cdot \widetilde P_3 \; n \cdot \wp}{\left[ n \cdot (\widetilde P_2 + \widetilde P_3) \right]^2} \;
\vep(\widetilde P_2) \cdot \vep(\widetilde P_3) \;\; \vep(\widetilde P_1) \cdot \vep^{*}(\wp)
\nn \\
&& \Bigl.
 + \frac{n \cdot \widetilde P_3  \; n \cdot \widetilde P_2 
- n \cdot \widetilde P_1 \; n \cdot \wp}{\left[ n \cdot (\widetilde P_2 + \widetilde P_1) \right]^2} \;
\vep(\widetilde P_2) \cdot \vep(\widetilde P_1) \;\; \vep(\widetilde P_3) \cdot \vep^{*}(\wp)
\;\; \Bigr\} \;\;,
\eeeq
where $g_S$ is the QCD coupling constant.
Note that the (physical) polarization vectors $\vep(\wp_i)$ 
and $\vep(\wp)$ in Eqs.~(\ref{colord3g}) and
(\ref{colord4g}) are defined in tha axial gauge with $\vep(p) \cdot n = 0$, where $n^\mu$
is the auxiliary vector introduced to specify the collinear limit 
(see Eq.~(\ref{ptilm})). Therefore both $V^{(3)}$ and $V^{(4)}$ depend on $n^\mu$ through
$\vep$.
The four-gluon effective vertex has an additional dependence on $n^\mu$ through
the momentum fractions $n \cdot\wp_i/n \cdot\wp_j$.

The recursion relation in Eq.~(\ref{eq:generalgluon}) is an equation of the 
Schwinger--Dyson type, and it is similar to the Berends--Giele recursion relation
\cite{Berends:1987me}
(see also Ref.~\cite{Draggiotis:1998gr})
for the (colour-ordered) multigluon off-shell current $J^\mu(1,\dots,m)$.
Note, however, that the splitting amplitudes are on-shell quantities, and the 
effective vertices $V^{(3)}$ and $V^{(4)}$ in Eq.~(\ref{eq:generalgluon}) are also
{\em on-shell} quantities (the Berends--Giele recursion relation uses the customary
three-gluon and four-gluon QCD vertices). Indeed, these vertices are fully specified
(see Eqs.~(\ref{colord3g}) and (\ref{colord4g})) by on-shell (light-like) parton momenta
$\wp_i$ and their corresponding on-shell (physical) polarization vectors 
$\vep(\wp_i)$.
This on-shell character of Eq.~(\ref{eq:generalgluon}) makes it somehow analogous
to the BCFW recursion relations
\cite{Britto:2004ap},
which directly construct on-shell amplitudes by joining on-shell amplitudes (with
lower multiplicity) through scalar propagators. 

The on-shell features of the recursion relation in Eq.~(\ref{eq:generalgluon})
are more evident by proceedings as follows. Using Eq.~(\ref{eq:generalgluon})
with $m=2$, we obtain $\s(1,2;\wp_{1,2})$ in terms of $V^{(3)}$ and the scalar propagator
$1/s_{1,2}$. This relation can be inverted to express $V^{(3)}$ in terms of 
$\s(1,2;\wp_{1,2})$.
Then, using Eq.~(\ref{eq:generalgluon})
with $m=3$, we obtain $\s(1,2,3;\wp_{1,3})$ in terms of $V^{(4)}$ and
$\s(i,j;\wp_{i,j})$ (i.e., $V^{(3)}$ and scalar propagators). This relation
can be inverted to express $V^{(4)}$ in terms of $\s(1,2,3;\wp_{1,3})$
and $\s(i,j;\wp_{i,j})$.
This implies that 
$\s(1,\ldots,m;\wp_{1,m})$ with $m \geq 4$ can in turn be {\em entirely} expressed in terms 
of scalar propagators and the splitting amplitudes with $m=2$ and $m=3$ gluons.
In summary, the recursion relation in Eq.~(\ref{eq:generalgluon}) gives the explicit result
for the splitting amplitudes with $m=2$ and $m=3$ gluons and, then, 
using these two splitting amplitudes as building blocks,
the same relation iteratively gives the explicit result for the
splitting amplitude with an arbitrarily-large number of collinear gluons.  

\section{Summary \& Outlook}

In Ref.~\cite{Catani:2012tb}
we have studied the multiparton collinear limit of generic tree-level scattering amplitudes
by using the (process-independent) splitting matrix formalism 
\cite{Catani:2003vu}.
We have derived recursion relations for the splitting functions that determine the
singular behaviour of the multiparton collinear limit for an arbitrary number of
gluons, quarks and antiquarks. 
The recursion relations display a self-organized structure based
on 2-parton and 3-parton building blocks 
(splitting functions or, equivalently, effective vertices)
for all the steps of the recursion.
The recursion relations, their derivation and applications are presented in a forthcoming
paper \cite{Catani:2012tb}.
In Sect.~\ref{sec:recrel} of this contribution, we have anticipated and presented 
the recursion relation for the pure multigluon case.
 
\section*{Acknowlegdements}

This work is supported by REA Grant Agreement PITN-GA-2010-264564 (LHCPhenoNet), by the MICINN
Grant No. FPA2007-60323, and FPA2011-23778, by CPAN (Grant No. CSD2007-00042),
by the Generalitat Valenciana Grant No. PROMETEO/2008/069, and by
INFN-MICINN agreement AIC-D-2011-0715.

\end{document}